\documentclass[aps,floatfix,epsfig]{revtex4}
\usepackage{pdfpages}
\usepackage{epsf}
\usepackage{amsmath}
\usepackage{eqnarray,amsmath}
\usepackage{epstopdf}
\usepackage{mathrsfs}
\usepackage{natbib}
\usepackage{amssymb,latexsym}
\usepackage{dcolumn}
\usepackage{graphicx}

\def\be{\begin{equation}}
\def\ee{\end{equation}}
\def\ba{\begin{eqnarray}}
\def\ea{\end{eqnarray}}

\graphicspath{{images/}}
\begin{document}
\title{\large \bf  Long gradient mode and large-scale structure observables I: linear order }
\author{Alireza Allahyari}
\affiliation{Department of Physics, Sharif University of Technology,
	Tehran, Iran }
\email{allahyari@physics.sharif.edu}

\author{Javad T. Firouzjaee}
\affiliation{ School of Astronomy, Institute for Research in Fundamental Sciences (IPM), P. O. Box 19395-5531, Tehran, Iran }
\email{j.taghizadeh.f@ipm.ir}

\begin{abstract}
We study the effect of long gradient modes on large scale observables. When defined correctly, genuine observables should not only be gauge invariant but also devoid of any gauge artifacts. One such gauge artifact is a pure gradient mode. Using the relativistic formulation of large scale observables, we confirm that a long gradient mode which is still outside observer's horizon leaves no imprint on the large scale observables at first order. These include the cosmic rulers and the number counts. This confirms the existing method for relativistically defined observables. The general relativistic bias relation for the halos and galaxies is also invariant under the presence of a long gradient mode perturbation. The observed power spectrum is not affected by this long mode.
\end{abstract}
%
%
\maketitle
\section{introduction}

Latest developments in the cosmology are provided by deep redshift surveys for cosmological structures which are the result of the evolution of cosmological perturbations
generated during inflation. Quantum fluctuations of the inflation field set the seeds of curvature perturbations at primordial epochs. The nature of these fluctuations are well studied in Cosmic Microwave Background (CMB) which is closely related to the inflationary early universe model. The fingerprints of these fluctuations at large scale structures (LSS) can be interesting since the current and future galaxy surveys will give high precision measurements of galaxy clustering, providing big statistical power to solve the issues in the standard model of cosmology since LSS surveys are three dimensional whereas CMB is two dimensional. \\

In galaxy surveys measurements the light propagates at the finite speed through the Universe and is modified by the inhomogeneity and the curvature of the matter. Therefore, we need to use proper general relativistic calculations to relate the observables which we observe from the light to the
physical quantities of source galaxies and the inhomogeneities properties that change the photon propagation. Since the exact relativistic calculation for cosmological structure observables has complexity \cite{exactsolution}, usually first or second order perturbation of the Einstein equation are used to calculate these observables. 
Some of these effects to the observed fluctuations of galaxies are the dark matter density fluctuations, the redshift-space distortions (peculiar velocity effect in the redshift space of light) and the magnification bias where studied in the \cite{ Jeong:2014ufa, Yoo:09-10, Bonvin:11-14, Challinor:2011bk}.
The standard way for calculating the full relativistic formula of galaxy clustering is done with the relation to the inhomogeneities and the source galaxy population, by tracing back the photon path given the observed redshift and the angular position of the source galaxies. Jeong et al. (for review see \cite{Jeong:2014ufa}) obtained a fully general  relativist expression for the observed galaxy density contrast at linear order, a fundamental galaxy clustering observable, including the
volume distortions due to the light deflection, evolving number density, galaxy density bias, as well as the magnification bias generalized to evolving luminosity function.\\

Generally, there are different wavelengths for the gravitational potential which come from the early universe fluctuations. The long wavelength potential can produce a superhorizon perturbation that might have effects on the CMB map or in the large scale structure observables. The power asymmetry in CMB map as observed by Planck satellite \cite{planck13} (for earlier reports of hemispherical asymmetry in WMAP data see \cite{wmap}) can be the generic property of the early universe model. The observations show that the power spectrum in northern hemisphere is different than the the power spectrum in southern hemisphere. Erickcek et al. \cite{Erickcek:2008sm} have proposed a superhorizon perturbation would introduce a preferred direction that generates the power asymmetry. Along this way, the predictions of inflationary models with long mode modulation of large scale structures are presented in \cite{Namjoo:2014nra}. As a matter of fact, it was shown that a long constant \cite{Weinberg} and a long gradient mode \cite{Hinterbichler:2012nm} can be gauge artifacts of the perturbation theory which does not leave any effect  on the cosmological observables \cite{Creminelli:2011sq, Mirbabayi:2014hda}. Our goal in this paper is to investigate the long gradient mode effects on the large scale observables  in the relativistic formulation.


The plan of the paper is as follows. In section II we will review the long mode effect in the perturbation theory of the large scale structure. Then, We calculate the three large-scale observable in the presence of long gradient mode in section III. The section IV is devoted to investigating the galaxy clustering quantities such as galaxy number counts and halo bias in the presence of the long gradient mode. Finally, we conclude in section V. The Latin indices indicate the space components and Greek indices indicate the space time components and we have set $c=1$.
\section{long gradient mode}
The perturbation theory defines gauge invariant quantities on spatial hypersurfaces with the assumption that perturbations fall of at the infinity. This allows to decompose perturbations as scalar, vector, and tensor (SVT). In the perturbation theory zero momentum modes will modulate the power spectrum in case of the single field slow roll inflation as derived in \cite{Maldacena:2002vr}, called the consistency relations. It is argued that the consistency relations are true in models of inflation in which the only dynamical field is the inflaton field \cite{Creminelli:2004yq}. This will induce a local type non-Gaussianity. Although this type of non-Gassainity is small and proportional to $n_{s}-1$, this is not clear how physics outside our horizon can  influence the local physics. Its is crucial to know if our large scale observations are really contaminated by such effects.
 
It has been shown that even after fixing the gauge  zero momentum transformations are allowed \cite{Weinberg}. This remaining gauge freedom is shown to be the source of IR divergences \cite{Urakawa:2010it}. This is attributed to the fact that we will need boundary conditions to uniquely solve for the lapse function after imposing the conventional comoving gauge condition. One may also say that constant modes can be removed by a coordinate transformation, so they should have no observable effect as they induce a relative shift in the expansion history. Let us suppose that we have a long mode perturbation that in a gradient expansion only the term $\frac{k}{a H}$ is dominant, where $a$, $H$ and $k$ are scale factor, Hubble parameter and Fourier mode for the long mode respectively. 
This potential could appear in an expansion of the primordial perturbation of the form $\sin k.x $. The average of such perturbations  should vanish in the whole universe in order to keep homogeneity and isotropy. We suppose that the effect of such perturbation does not vanish in our Hubble patch.
Such a perturbation does not fall off at infinity as required by the perturbation theory. One may expect that such a pure gradient mode in single field slow roll inflations must have no observable imprint on the cosmological observables, since as the long mode freezes outside the Hubble radius it turns into an adiabatic mode. The effect of a pure gradient mode is then an acceleration which is not observable locally.\\
This seems to be justified in the separate universe approach in which the universe can be locally approximated by a curved FRW on patches smaller than the long mode as long as we are not measuring correlation between patches \cite{Wands:2000dp}. Here this patch is well beyond our observable universe.
The physical effects of a long mode start at $(\frac{k}{aH})^{2}$ order, giving rise to an effective local curvature in the case of scalar perturbations. Long tensor modes at order $(\frac{k}{aH})^{2}$ will imprint a quadrapole in density fluctuations on small scale structures before tensor modes are damped away \cite{Dai:2013kra}.
By extension of Weinberg adiabatic modes it is shown that a pure gradient mode can be removed from our entire horizon \cite{Hinterbichler:2012nm}. We may require that true observables are invariant under both normalizable and non-normalizable transformations \cite{Urakawa}.
The vanishing effect of a constant mode in the power spectrum is emphasized in \cite{Dai:2013kra}. In this case it is crucial to note that the power spectrum should be scale invariant. For the CMB, we can see this vanishing contribution if change of last scattering surface position is taken into account \cite{Creminelli:2011sq}.

Now we are interested in the case of a pure gradient mode.
Using the conformal Fermi coordinates one can show that time independent gradient modes are removed in patches smaller than the gradient mode wavelength and the tidal effects vanish identically \cite{Dai:2015rda}. We are justified to say that pure gradient modes leave no imprint on dynamics locally. However, because we observe on our past light cone, gradient modes appear in relations between the physical quantities and apparent observed ones. Thus, the vanishing effect of such long modes in observables is a consistency  check of the proposed formulation for large scale observables.

To this end, we need to define observables. Observables should be gauge invariant as a priori and made of information on our past light cone. They should be devoid of gauge artifacts when written in any gauge.
They are described by the standard clocks, the standard rulers and the number counts \cite{Jeong:2014ufa}. One example of such a clock is the CMB on large scales.
In the case of the CMB it is shown that a gradient mode leaves no imprint at first order using suitable coordinate transformations and averages when temperature is physically defined for the CMB and the observer velocity is taken into account\cite{Mirbabayi:2014hda}. A consistent general relativistic formulation of galaxy clustering should naturally take into account effects such as displacements, velocities and other new relativistic effects.  
The other cosmological observable is galaxy number counts. Future surveys will probe cosmos on large volumes with unprecedented precision. Because of their large statistical power, they will be able to discriminate between inflationary models by the level of non-Gaussianity. It is crucial to consistently formulate all effects in order to  correctly interpret the data.
A fully relativistic formulation has recently emerged \cite{Jeong:2014ufa, Yoo:2014kpa, Bonvin:11-14}. The authors in \cite{Challinor:2011bk} use the evolution equation for the Jacobi map to derive the number counts. As a consistency check we show that a pure gradient mode must have no observable imprint on observed galaxy number counts for comoving sources and observers.
\section{large-scale observables}
Kaiser formula defines the relation between the observed power spectrum in the redshift space and the physical power spectrum \cite{Kaiser}. However, for large sky surveys this formula is hindered by the fact that this formula does not consistently take into account all relativistic effects. In addition, on large scales, gauge effects become important which should be taken into account and we will need a gauge invariant formulation. For example galaxies are thought to trace the underlying dark matter overdensity and are biased by a scale independent factor $b_{g}$ with respect to the dark matter overdensity. This definition is obscured because changing the gauge will introduce scale dependencies. We also need to define observables which should  be gauge invariant in prior and physically defined as they will be explained. Since photons are traveling through inhomogeneities, their path will  deviate and these inhomogeneities change the observed size of our rulers and our apparent survey volume.In this section we investigate the photon propagation effects from inhomogeneities.

An observer sees  the cosmos images by projecting on his own instantaneous screen space \cite{ellis book} spanned by orthonormal basis $\textbf{e}_{1}$ and $\textbf{e}_{2}$ which are orthogonal to the observer's velocity. His velocity defines instantaneous volume in which he measures.
When a photon is observed at a redshift $z$ (Observed redshift), its apparent position is attributed as $\tilde{x}^{\mu}(z)$
\footnote{$\tilde{} $ refers to apparent quantities.}.
The photon trajectory at the observed redshift $z$ is deviated from a similar trajectory in a homogeneous universe. The actual position of a source is obtained by integrating the geodesic equations for photons. These deviations are gauge dependent and unobservable.
Because null geodesics are conformally invariant we can use conformally transformed metric and affine parameter for the photons. The source position is given by $$x^{\mu}=\bar{x}^{\mu}(z)+\Delta x^{\mu}=\bar{x}^{\mu}(z)+\delta x^{\mu}+\frac{d{\bar{x}}^{\mu}}{d\chi}\delta\chi,$$ where $\chi$ is the affine parameter perturbation and $\bar{x}^{\mu}(z)$ is the background position evaluated at the observed redshift.  The first term is given by the geodesic equation and the second term is given by the redshift matching as it will be explained. The perturbed conformal photon four-vector is given by $k^{\mu}=(-1+\delta \nu, n^{i}+\delta n^{i})$. When the observer fixes the scale factor at  his observation time $a(t_{p})=1$, the global scale factor will be $a(\tau_{o})\neq 1$ where $t_{p}=\int_{0}^{\tau_{0}} \sqrt{-g_{00}(x_{o},\tau)}a(\tau)d\tau$ is the observer's proper time and the subscript $o$ refers to the observer. The metric in a general gauge is written as $$ds^2=a^2(-(1+2A)d\tau^2-2B_{i}d\tau dx^{i}+(\delta_{ij}+h_{ij})dx^i dx^{j}).$$
We work in the conformal Newtonian gauge in which the metric is given by
\begin{align}
ds^2=a^2(-(1+2\varphi)d\tau^2+(1-2\psi)dx^i dx_{i}).
\end{align}
In the matter dominated Einstein-de Sitter (EdS) universe $\varphi=\psi$.
The scale factor difference between the global scale factor $a(\tau_{o})$ and the $a(t_{p})$ is $\delta a=a(\tau_{o})-a(t_{p})=-H_{o}\int_{0}^{\tau_{o}}\varphi(x_{o},\tau) a(\tau) d\tau$.
To solve the geodesic equations we need the initial conditions which are set by requiring that in observer's frame the frequency and direction of the photons are given by
\begin{align}
1=(a^{-2}g_{\mu\nu}e^{\mu}_{0} k^{\nu})_{o}\\ \nonumber
n_{i}=(a^{-2}g_{\mu\nu}e^{\mu}_{i}k^{\nu})_{o},
\end{align}
where $e^{\mu}_{\nu}$ are orthonormal tetrads given by $e^{\mu}_{0}=u^{\mu}=a^{-1}(1-\varphi,v^i)$ and $e^{u}_{i}=(v_{i},\delta_{i}^{j}+\psi\delta_{i}^{j})$. These will fix the perturbation at observer's position as
\begin{align}
\delta\nu_{o}=-\delta a+\varphi_{o}+v_{\lVert}\nonumber\\
\delta n^i_{o}=n^{i}\delta a-v^i_{o}+\psi n^i,
\end{align}
\footnote{For any vector $v^{i}$ we have defined $v_{\lVert}=v^{i}n_{i}$ and $v^{i}_{\perp}=\mathcal{P}^{ij}v_{j}$ in which $\mathcal{P}^{ij}=\delta^{ij}-n^{i}n^{j}$ is the projection operator.}.
Integrating geodesic equation  and using the initial conditions give
\begin{align}
\delta x^{0}=(-\delta a-\varphi_{o}-v_{||o} )\chi+\int \left(  2\varphi+(\chi-\chi^{\prime})(\dot{\varphi}+\dot{\psi}) \right)d\chi^{\prime}-\int_{0}^{\tau_{o}} \varphi(x_{o},\tau)ad\tau \\
\delta x^i=(\delta a n^i-\psi_{o}+v_{o}^i)\chi +\int \left(\psi n^i+(\chi-\chi^{\prime})(-\partial_{i}\varphi-\partial_{i}\psi) \right)d\chi^{\prime}. 
\end{align}
On the other side, the redshift for the photons is defined by
\begin{align}
1+z\equiv \frac{1}{\tilde{a}}=\frac{(k_{\mu}u^{\mu})_{e}}{(k_{\mu}u^{\mu})_{o}}=\frac{(1+\varphi+v_{\lVert}-\delta\nu)_{o}}{a(x^{0})}.
\label{redshift}
\end{align}
The scale factor perturbation with respect to the apparent scale factor $\tilde{a}$ is defined as $\Delta \ln a=\frac{a}{\tilde{a}}-1$. We can write this as 
\begin{align}
\Delta \ln a=\frac{\partial \ln a}{\partial \tau}(x^{0}-\tilde{x}^{0}).
\end{align}
Using (\ref{redshift}) we get 
\begin{align}
\delta\chi=\delta x^{0}-\frac{1+z}{H}\Delta \ln a.
\end{align}

\subsection{Cosmic rulers}
Cosmic rulers are cosmological observables that their spatial scale $r_{0}$ is known like CMB on large scales or a known scale in the matter power  spectrum  like baryon acoustic oscillation (BAO) \cite{Fabian}.  Their observed value is related to the cosmic candle perturbations which are related to luminosity distance perturbations. These are used for studying the nature of dark energy \cite{Barausse:2005nf}.
Cosmic ruler observed scale changes from $r_{0}$ to the observed apparent scale $\tilde{r}$. Note that the size of the rulers may change with time by a term proportional to cosmic clock perturbations. The cosmic clock perturbations are not affected by pure gradient modes as we will explain. Hence we consider non evolving rulers. To define their length we need to define the observer.
Their length is defined as the length measured in instantaneous frame of the comoving observers whose velocity is given by $u^i=\frac{T^{i}_{0}}{\rho+p}$ and his projected metric is given by $(g_{\mu\nu}+u_{\mu}u_{\nu})$. If the size of the ruler is small we can approximate its apparent length as
\begin{align}
\tilde{r}^{2}=\tilde{a}^{2}(z)\left( -(\delta \tilde{x}^{0})^2+\delta_{ij}\delta\tilde{x}^{i}\delta\tilde{x}^{j}\right), 
\end{align}
whereas its physical scale is given by
\begin{align}
r_{0}^{2}=(g_{\mu\nu}+u_{\mu}u_{\nu})\left(\delta\tilde{x}^{\mu}+\Delta x^{\mu}-\Delta x^{\prime \mu} \right)\left( \delta\tilde{x}^{\nu}+\Delta x^{\nu}-\Delta x^{\prime \nu}\right),  
\end{align}
where $\delta \tilde{x}^{u}=x^{\mu}-x^{\prime \mu}.$  
At first order using again the small ruler approximation, $\Delta x^{i}-\Delta x^{\prime i}\simeq \delta \tilde{x}^{\beta}\frac{\partial}{\partial \tilde{x}^\beta}\Delta x^{i}$, and small angel approximation, $\delta\tilde{x}^{0}=-\delta\tilde{x}_{\lVert}$, 
the relative ruler perturbation is
\begin{align}
\frac{\tilde{r}-r_{0}}{\tilde{r}}=\mathcal{C}\frac{(\delta \tilde{x}_{\lVert})^2}{\tilde{r}^{2}_{c}}+\mathcal{B}_{i}\frac{\delta \tilde{x}_{\lVert}\delta \tilde{x}^{i}_{\perp}}{\tilde{r}_{c}^{2}}+\mathcal{A}_{ij}\frac{\delta \tilde{x}^{i}_{\perp}\delta \tilde{x}^{j}_{\perp}}{\tilde{r}^{2}_{c}},
\end{align}
where $\tilde{r}_{c}\equiv \frac{\tilde{r}}{a}$ \cite{Fabian}.
First, we will introduce each of these geometric factors and then we calculate the effect of a pure gradient mode on these observables.

\subsubsection{2-scalar C}
The function $\mathcal{C}$  gives the line of sight perturbations of cosmic rulers and causes perturbations in redshift space. It also encompasses the redshift space distortion term which is the dominant term on small scales. In the case of non-evolving rulers, the function $\mathcal{C}$  is given by
\begin{align}
\mathcal{C}=-\Delta \ln a-\frac{1}{2}h_{\lVert}-v_{\lVert}-\partial_{\chi}\Delta x\lVert, 
\end{align}
where $h_{\lVert}=h_{ij}n^{i}n^{j}$.
The $\mathcal{C}$ function should be devoid of any gauge artifact since it is an observable. To check the formulation let us calculate the effect of a pure gradient mode on  $\mathcal{C}$. In the EdS universe a pure gradient mode is given by $\psi=\phi=k.x$ where $\frac{k}{aH}< 1 $. In the Newtonian gauge  $\mathcal{C}$  is given by
\begin{align}
\mathcal{C}=-\Delta\ln a \left( 1-H(z)\frac{\partial}{\partial_{z}}(\frac{1+z}{H})\right)-\varphi-v_{\lVert}+\frac{1+z}{H(z)}(-\partial_{\lVert}\varphi-\dot{v}_{\lVert})=\left( +\frac{2}{3}\chi(\frac{3}{2})-\chi+\sqrt{a}(\frac{1}{H_{o}}-\frac{3}{2})\right) k_{\lVert}=0,
\end{align}
where $v_{\lVert}=-\frac{2}{3}\frac{\sqrt{a}k_{\lVert}}{H_{o}}$, $k_{\lVert}=k_{i}n^{i}$ and $\chi=\frac{2}{H_{o}}(1-\sqrt{a}).$ We used the fact that $\partial_{\lVert} v^{i}=n^{j}\partial_{j} v^{i}=0$.
We also have $$
\Delta \ln a=v_{\lVert}-v_{o}-\varphi=-\frac{2}{3}\chi k_{\lVert}.$$
Note that neglecting first order perturbations we have $H_{o}=\frac{3}{2}$.
As a result, the line of site perturbations are not affected by the presence of the pure gradient mode.

\subsubsection{2-vector $\mathcal{B}$}
The vectorial term which is a projected vector can be written as
\begin{align}
\mathcal{B}_{i}=-v_{\perp i}+\frac{1+z}{H(z)}\partial_{\perp i}\Delta\ln a,
\end{align}
where $\partial_{\perp i}={P}_{i}^{j}\partial_{j}=(\delta_{i}^{j}-n_{i}n^{j})\partial_{j}$.
This term produces perturbations both in the line of sight and perpendicular to the line of sight components.
Any spin one quantity such as  $\mathcal{B}_{i}$ term can be written in the spin basis as
\begin{align}
_{\pm1}\mathcal{B}=m^{i}_{\mp}\mathcal{B}_{i}=-v_{\pm}+\frac{1+z}{H}\partial_{\pm}\Delta \ln a
\end{align}
 where $\textbf{m}_{\pm}=\frac{\textbf{e}_{1}\mp i\textbf{e}_{2}}{\sqrt{2}}$ and $\textbf{e}_{1}$ and $\textbf{e}_{2}$ are vector basis on the sphere. For a gradient perturbation we have $\mathcal{B}_{i}=(\frac{2}{3H_{0}}-1)\sqrt{a}k_{\perp i}=0$.
Again a pure gradient gives vanishing contribution to $\mathcal{B}_{i}$. Similar to the former case, the $\mathcal{B}$ term is not changed by the pure gradient mode.
\subsubsection{Magnification and shear}
The most crucial term is $\mathcal{A}_{ij}$ which is a transverse tensor which can be written as 
\begin{align}
\mathcal{A}_{ij}=-\Delta \ln a\mathcal{P}^{i}_{j}-\frac{1}{2}\mathcal{P}_{i}^{k}\mathcal{P}_{j}^{l}h_{kl}-\partial_{\perp ( i}\Delta x_{\perp j)}-\frac{1}{\tilde{\chi}}\Delta x_{\lVert}\mathcal{P}_{ij}.
\end{align}
Magnification which is the trace part of $\mathcal{A}_{ij}$  is $$\mathcal{M}=\mathcal{P}^{ij}\mathcal{A}_{ij}=-2\Delta \ln a -\frac{1}{2}(-4\varphi)+2\hat{k}-\frac{2}{\chi}\Delta x_{\lVert}$$ where $\hat{k}=-\frac{1}{2}\partial_{\perp i}\Delta x^{i}_{\perp}$. Note that lensing convergence ($\hat{k}$) is not gauge invariant. Hence, it is not observable. It has been shown that magnification is not perturbed by gradient modes \cite{Fabian}. The trace free part of the $\mathcal{A}_{ij}$ term is called shear, $\gamma _{ij}$,  which is a spin two quantity. The shear term can be written in the spin basis as $_{\pm2}\gamma=m^{i}_{\mp}m^{j}_{\mp}\gamma_{ij}$.
In the Newtonian gauge in the spin basis the shear term  is given by
\begin{align}
_{\pm2}\gamma=\int(\tilde{\chi}-\chi)\frac{\chi}{\tilde{\chi}}m^{i}_{\mp}m^{j}_{\mp}\partial_{i}\partial_{j}(2\varphi)d\chi.
\end{align}
Since the shear is related to the second derivatives of the metric, it is not affected by gradient modes. Consequently surveys like weak lensing surveys which measure the shear are not affected  by the gradient modes.

\subsubsection{Luminosity distance}
The magnification which produces area perturbations is measured in the lensing surveys and surveys which probe the luminosity like supernova surveys. It is related to angular and luminosity distance by $\frac{\Delta D_{A}}{D_{A}}=\frac{\Delta D_{L}}{D_{L}}=-\frac{1}{2}\mathcal{M}$ \cite{Bonvin:2005ps}.
Therefore, a pure gradient should induce no asymmetry in our cosmic candle observations.

\section{galaxy number counts}
One of the most important observable quantities in the large scale structure surveys is the galaxy number counts. We are interested in the effect of the long gradient modes in the number count surveys. Since we observe tracers and not the underlying matter field we will need to define a general relativistic bias relation applicable for long modes. It is argued that the bias relation in the synchronous gauge is appropriate \cite{Donghui}. This is in contrast to the advocated gauge  in \cite{Yoo:09-10} that chooses constant redshift gauge for the bias relation.  The synchronous gauge is proposed for a second order calculation of the galaxy clustering \cite{Yoo:2014vta}. 
Suppose we have a comoving source with four velocity $ u^\mu$. In the instantaneous source rest frame a volume element is defined by $dV_{\mu}=\varepsilon_{\mu\nu\alpha\beta}dx^{\nu}dx^{\alpha}dx^{\beta}$. The number of galaxies defined in the observed coordinates is given by
\begin{align}
N(x)=\int\sqrt{-g} n_{g} u^{\mu}\varepsilon_{\mu\nu\alpha\beta}\frac{\partial x^{\nu}}{\partial{{\tilde{x}}^{1}}}\frac{\partial x^{\alpha}}{\partial{{\tilde{x}}^{2}}}\frac{\partial x^{\beta}}{\partial{{\tilde{x}}^{3}}}d^3{\tilde{x}}. 
\end{align}
where $n_{g}$ is the physical number density of the galaxies.
The number of the galaxies can be written as
\begin{align}
N(x)=\int (1+\varphi-3\psi)a^{3}\bar{n}_{g}(z)(1+\delta_{g}(z,x))[(1-\varphi)\arrowvert\frac{\partial x^i}{\partial \tilde{x}^j}\arrowvert+v_{\lVert}]d^{3}\tilde x=\int \tilde{a}^{3}\tilde{n}_{g}(z,\tilde{x})d^{3}\tilde{x},
\end{align}
where $\delta_{g}$ is number density perturbation and $x^{i}=\tilde{x}^{i}(z)+\Delta x^{i}$.
Thus, the observed number of the galaxies in the observed coordinates is given by 
\begin{align}
\delta_{g}(z,\tilde{x})=\delta_{g}(z,x)+\delta V=\delta_{g}-\psi+v_{\lVert}+\partial_{\lVert}\Delta x_{\lVert}+\frac{2}{\chi}\Delta x_{\lVert}-2\hat{k}
\label{galaxy-p},
\end{align}
where $\hat{k}=-\frac{1}{2}\partial_{\perp i}\Delta x^{i}_{\perp}$. This formula is the relativistic generalization of Kaiser formula which includes the new relativistic corrections. These corrections will become important on the large scales where general relativity and other models may deviate from each other. These corrections should be considered in the future surveys. On these scales we are constrained by the cosmic variance. Detectability of these new relativistic corrections is discussed in \cite{Yoo-test}. The method proposed to overcome cosmic variance is given by multi tracer method which uses the fact that different biased tracers trace the same underlying density field \cite{Seljak}.

We check how a pure gradient mode affects the
$\delta V$ in the relativistic formulation.
Consider the case of a matter dominated universe with sources and the observers comoving with the cosmic fluid. In the presence of the long mode we have $v^i=-\frac{2}{3}\frac{\sqrt{a}k^i}{H_{o}}$ and $v_{o}^{i}=-\frac{2}{3}\frac{1}{H_{o}}$.
In the Newtonian gauge each term in (\ref{galaxy-p}) is given by
\begin{align}
\partial_{\lVert}\Delta x_{\lVert}=\frac{5}{3}\chi-\frac{3}{2}\sqrt{a}(-1+\frac{2}{3H_{o}}),\\
\frac{2}{\chi}\Delta x _{\lVert}=2\chi+2\sqrt{a}
k_{\lVert}=(\frac{2}{3H_{o}}-\chi)k_{\lVert}.
\end{align}	
Using (\ref{galaxy-p}) we find that 
\begin{align}
\delta V=0.
\end{align}
Consequently the number count observations are not changed by this long mode.
Since we observe galaxies and not the underlying dark matter density field, we need a bias relation to relate $\delta_{g}^{t_{p}}$ in the synchronous gauge to the dark matter perturbation $\delta^{t_{p}}$. As we observe on constant redshift surfaces, this bias relation picks up a perturbation given by $ \delta_{g}^{t_{p}}=b\delta^{t_{p}}+\frac{d \ln n_{g}}{d \ln a}\delta \ln a=b\delta^{t_{p}}+\frac{d \ln n_{g}}{d \ln a}\mathcal{T}(\boldmath n)$. The $\mathcal{T}(\boldmath n)$ is the cosmic clock perturbation \cite{Jeong:2013psa}. The cosmic clock perturbation by the gradient mode vanishes as 
\begin{align}
\mathcal{T}(\boldmath n)=-\frac{1}{3} \varphi+v_{\lVert}-v_{o}=0,
\end{align}
where it is assumed that observers are comoving with the cosmic fluid $v_{\lVert}=-\frac{2}{3}\frac{\sqrt{a}k_{\lVert}}{H_{o}}$.

\subsection{Halo bias}
The number density of halos, $n_{h}$, is biased by the presence of long mode perturbations. In Press-Schechter method number density of small scale perturbations in presence of long mode perturbations $(\nu_{b})$ is 
\begin{align}
\mathcal{N}_{pk}(\nu_{s}|\nu_{b})d\nu_{s}=\frac{\mathcal{N}_{pk}(\nu_{s},\nu_{b})d\nu_{s}d\nu_{b}}{\mathcal{P}(\nu_{b})d\nu_{b}}
\label{bias},
\end{align}
where $\nu=\frac{\delta}{\sigma}$ in which $\sigma$ is the variance of matter fluctuations, $\mathcal{N}_{pk}(\nu_{s},\nu_{b})$ is the number density of peaks where the long mode has an amplitude $\nu_{b}\pm d\nu_{b}/2$ and $\mathcal{P}$ is the distribution function. Long mode perturbations locally act as a background. For long mode fluctuations equation (\ref{bias}) simplifies to  $$\mathcal{N}_{pk}(\nu_{s}|\nu_{b})\backsimeq\mathcal{N}_{pk}(\nu_{p})$$ where $\nu_{p}=\frac{\delta_{s}-\delta_{b}}{\sigma_{s}}$ \cite{simon}. This will lead to the bias relation for the fluctuations of peak number density given by  $\delta_{h}=b_{h}\delta_{b}$. This bias relation depends on $\sigma_{8}$ which is the matter variance at $8$ Mpc. In this formulation long modes change the threshold of the halo formation, $\delta_{c}$, which is obtained by the spherical collapse model to $\delta_{c}-\delta_{b}$ where $\delta_{b}$ is the long mode over density.
We are interested in the effect of the pure gradient modes on the halo bias. A long gradient mode does not change the bias relation $b_{h}$ because $\delta_{b}=0$. However, on large scales gauge effects become important. Changing the gauge will produce scale dependencies in the bias relation. Thus, one needs to find a relativistic definition for the bias relation. The synchronous gauge is assumed to be the right gauge. Since halo formation is a local phenomena, it only depends on the dark matter perturbations in this gauge. This gauge has been implemented for writing the relativistic galaxy clustering. The long gradient By going to the Fermi normal coordinates we can separate the local effects of long modes. These coordinates can capture local effects for patches smaller than the Hubble length but fail to describe the short mode dynamics when the initial conditions need to be imposed.
In this coordinate corrections start at quadratic order, $h_{ij}x^{i}x^{j}$, which produce the tidal effects. It is shown that for the long perturbations with spherical symmetry $h_{ij}$s are functions of second derivatives of the metric perturbations. The bias relation defined in the local Fermi normal coordinates depends on the second derivatives of the global metric and reduces to the usual bias relation on the small scales \cite{Tobias}. As a result number density of halos is not affected by pure gradient modes.

The other complexity is that we observe on the slices of constant redshift. This will produce another perturbation $$\delta_{h}(z)=b\delta^{t_{p}}+\frac{d \ln n_{h}}{d \ln a}\delta \ln a=b\delta_{t_{p}}+\frac{d \ln n_{h}}{d \ln a}\mathcal{T}(\boldmath n)$$ in the bias relation. As stated, cosmic clocks are not perturbed by pure gradient modes. Hence, the observed bias relation is not changed.
\subsection{Galaxy power spectrum}
Finally, we compute the effect of a pure gradient mode on the observed power spectrum. We can compute the intrinsic power spectrum in a chosen gauge and transform it back to the redshift space using the equations in \cite{Pajer:2013ana}. This relation is given by
\begin{align}
\tilde{\xi}(\tilde{r},z)=\left(1-a_{ij}\tilde{x}^{i}\partial^{j}+\mathcal{T}\partial_{\tilde{z}} \right) \xi(\tilde{r},\tilde{\tau}),
\end{align}
where 
\begin{align}
a_{ij}=\mathcal{C}\hat{n}_{i}\hat{n}_{j}+\hat{n}_{(i}\mathcal{P}_{j)k}\mathcal{B}^{k}+\mathcal{P}_{ik}\mathcal{P}_{jl}\mathcal{A}^{kl}.
\end{align}
since $\mathcal{C},\mathcal{B}^{k}, \mathcal{A}^{kl}$ and $\mathcal{T}$ do not change by the pure gradient presence, the observed large scale power spectrum is not affected.

\section{discussion and conclusion}

The vanishing effect of gradient modes can not be deducted from the equivalence principle.
It is shown that the local physical effects of  long mode perturbations start at $(k_{L}x)^2$ ( called tidal terms) if one uses conformal Fermi coordinates as implied by the equivalence principle \cite{Dai:2015rda}. As a result  the pure gradient mode does not contribute to the tidal term at linear order. However, to calculate the observables at late time, we have to transform from the conformal Fermi coordinates to the observed coordinates. Observed quantities change when they are mapped to the redshift space by terms which are proportional to the cosmic clocks and cosmic rulers \cite{Pajer:2013ana}. The equivalence principle does not imply that cosmic clocks and rulers are not affected by the presence of a pure gradient mode.\\

We have studied that the effect of a pure gradient mode on the large scale observables. First, we have shown that the contributions of gradient mode  to all cosmic ruler observables vanishes. This shows that the existing relativistic formulation of observables encompasses all the effects including observer's velocity and displacement. We have  shown that  different projection effects cancel each other in the relativistic galaxy clustering. Finally, we confirmed that the observed power spectrum does not change by the presence of the long mode.\\


{\bf Acknowledgments:}\\

I would like to thank Hassan Firouzjahi, Ali Akbar Abolhasani and Reza Mansouri for useful discussions and comments.\\

\end{document}